\documentclass[aps]{revtex4}
\usepackage{color}
\usepackage{psfrag}
\usepackage{dcolumn}
\usepackage{bm}
\usepackage[latin1]{inputenc}
\usepackage[spanish,english]{babel}
\usepackage{amsfonts}
\usepackage{amssymb,epsf}
\usepackage{graphicx}
\usepackage{epstopdf}
\usepackage{amsmath,amssymb}
\usepackage{pdflscape}
\usepackage{adjustbox}
\usepackage{hyperref}

\begin{document}

\title{Is the remnant of GW190425 a strange quark star?}
\author{J. Sedaghat$^{1}$\footnote{J.sedaghat@shirazu.ac.ir}, S. M. Zebarjad$^{1,2}$\footnote{zebarjad@shirazu.ac.ir}, G. H. Bordbar$^{1,3}$\footnote{ghbordbar@shirazu.ac.ir}, B. Eslam Panah$^{4,5,6}$\footnote{eslampanah@umz.ac.ir}, and R. Moradi$^{6,7}$\footnote{Rahim.Moradi@icranet.org},}
\affiliation{$^{1}$ Department of Physics, Shiraz University, Shiraz 71454, Iran\\
$^{2}$ Department of Physics, University of California at San Diego, La
Jolla, CA 92093, USA\\
$^{3}$ Department of Physics and Astronomy, University of Waterloo, 200
University Avenue West, Waterloo, Ontario N2L3G1, Canada\\
$^{4}$ Sciences Faculty, Department of Physics, University of Mazandaran, P.
O. Box 47415-416, Babolsar, Iran\\
$^{5}$ ICRANet-Mazandaran, University of Mazandaran, P. O. Box 47415-416,
Babolsar, Iran\\
$^{6}$ ICRANet, Piazza della Repubblica 10, I-65122 Pescara, Italy\\
$^{7}$ ICRA, Dipartimento di Fisica, Universit`a di Roma \textquotedblleft
La Sapienza\textquotedblright , Piazzale Aldo Moro 5, I-00185 Roma, Italy}

\begin{abstract}
This study investigates the effects of different QCD models on the structure of strange quark stars (SQS). In these models, the running coupling constant has a finite value in the infrared region of energy. By imposing some constraints on the strange quark matter (SQM) and exploiting the analytic and background perturbation theories, the equations of states for the SQM are obtained. Then, the properties of SQSs in general relativity are evaluated. By using component masses of GW190425 \cite{Abbott2020ApJL} as well as some conversion relations between the baryonic mass and the gravitational mass, the remnant mass of GW190425 is obtained. Our results for the maximum gravitational mass of SQS are then compared with the remnant mass of GW190425. The results indicate that the obtained maximum gravitational masses are comparable to the remnant mass of GW190425. Therefore, it is proposed that the remnant mass of GW190425 might be a SQS.
\end{abstract}

\maketitle

\section{Introduction}

Compact stars are considered large laboratories for investigating quantum chromodynamics (QCD) models. One of the most challenging issues in QCD is the running coupling constant's infrared (IR) scale behavior. The running coupling constant obtained from the renormalization procedure and the renormalization group equations has a well-defined behavior at large momenta \cite{Prosperi2007,Altarelli2013}. However, it becomes infinite at a
point in the IR scale called the Landau pole ($\Lambda $). Based on the perturbative
QCD, the quark confinement originates from the Landau pole, which depends on
the selected renormalization scheme \cite{Deur2016}. Nevertheless, the data
extracted from some experiments indicate that the running coupling constant
of QCD at low momenta is finite and freezes to a constant value \cite{Deur2016}. For example, the effective strong coupling constant ($\alpha
_{s,g1}(Q^{2})$) defined based on the Bjorken sum rule has been extracted
from the CLAS spectrometer \cite{Deur2008} where the polarized electron beam
(with energies ranging from $1$ to $6$$GeV$) collides with proton and
deuteron targets. The data show that $\alpha _{s,g1}(Q^{2})$ loses its scale
dependence at low momenta. In Ref. \cite{Perez-Ramos2010}, by analyzing the
energy spectra of heavy quark jets from $e^{+}e^{-}$ annihilation, the IR
value for the effective coupling constant ($\alpha _{s}^{eff}(Q)$), is
obtained as $\left( 2GeV\right) ^{-1}\int_{0}^{2GeV}(\alpha _{s}^{eff}(Q)/{\pi })dQ=0.18\pm 0.01$. Using the data of hadronic decays of the $\tau $%
-lepton extracted from the OPAL detector at LEP \cite{OPAL1999}, it is shown
that for a hypothetical $\tau $-lepton with the mass of $m_{\tau ^{^{\prime
}}}$, the effective charge $\alpha _{\tau }(m_{\tau ^{^{\prime }}}^{2})$,
freezes at the mass scale of $m_{\tau ^{^{\prime }}}^{2}\cong 1GeV^{2}$ with
the magnitude of $0.9\pm 0.1$ \cite{Brodsky2003}. Such behavior of the
coupling constant at low momenta is called IR freezing in the literature.
This effect can be explained by the running behavior of the coupling
constant, which stems from particle-antiparticle loop corrections. Due to
confinement, quarks and antiquarks cannot have a wavelength larger than the
size of the hadron. This suppresses the loop corrections at the IR scale.
Consequently, $\alpha _{s}$ is expected to lose its energy dependence at low
energies \cite{Brodsky2008}. In addition, the theoretical results from the
Lattice QCD \cite{Bali2001} and the Schwinger-Dyson framework \cite%
{Fischer2006}, show that $\alpha _{s}$ freezes at low momenta. There are
other models, such as the Stochastic quantization approach \cite%
{Zwanziger2002}, the optimized perturbation theory \cite{Mattingly1992}, the
Gribov-Zwanziger approach \cite{Gracey2006}, and the background perturbation
theory \cite{Badalian1997,Simonov2011}, in all of which $\alpha _{s}$ runs
with an IR freezing effect. Moreover, the analytic perturbation theory
presents a running coupling constant with a slowly varying behavior at low
energies \cite{Shirkov1997}. It should be noted that for the quark
confinement, the coupling constant does not need to be infinite in the IR
region. For example, the lattice simulations show that the $q\bar{q}$
potential increases linearly at distances larger than $0.4fm$, while the
coupling constant freezes at a maximum value \cite{Bali2006}. Contrary to
general belief, the value of the coupling constant does not need to be
infinite in the IR region in order to confine light quarks \cite{Gribov1999}.

This paper first investigates the IR behavior of the QCD running coupling constant in different models. Then, two models are selected for the perturbative calculation of the equations of states (EOSs) of strange quark matter (SQM) in the leading order of $\alpha _{s}$. These two models are i) the analytic perturbation theory (APT) and ii) the background perturbation theory (BPT). Afterward, the EOSs are used in the TOV equation to calculate the maximum gravitational masses of strange quark stars (SQSs). Motivated by the LIGO detection of the compact binary coalescence (GW190425) with the total mass of $3.4_{-0.1}^{+0.3}{M}_{\odot }$ \cite{Abbott2020ApJL}, we compared our results with the remnant mass of GW190425. This binary is
more massive than the other reported Galactic double neutron stars (NSs) \cite{Farrow2019}. Since no electromagnetic counterpart has been observed for GW190425, its origin is unknown \cite{Kyutoku2020}. Various models have suggested the nature of GW190425. In Ref. \cite{Kyutoku2020}, the possibility of whether GW190425 is a binary NS merger or a black hole-NS (BH-NS) merger has been investigated. It is hypothesized that the progenitor of GW190425 is a binary including a NS and a $4-5M_{\odot }$ helium star \cite{Romero-Shaw2020}. Furthermore, it is suggested that GW190425 is a NS-BH merger with the masses of $1.15_{-0.13}^{+0.15}{M}_{\odot }$ and $2.4_{-0.32}^{+0.36}{M}_{\odot }$ for the NS and the BH, respectively \cite
{Han2020}. In addition, using a toy model, some researchers have investigated whether future LISA observations could detect binary NSs like GW190425 \cite{Korol2021}. In Ref. \cite{Clesse2020}, GW190425 and GW190814 have been investigated as primordial BH clusters. In this paper, it is not intended to probe the nature of GW190425. Rather, the aim is to explore
whether the remnant mass of GW190425 is a SQS. In this study, it is shown that the maximum gravitational masses of SQSs are comparable with the remnant mass of GW190425. 

In this paper, we consider a pure SQS whose degrees of freedom are quarks (up, down, and strange flavors) and gluons. However, in a realistic description of SQS, the phase of the
hadronic matter (nucleons and hyperons) could occur at low densities at the star's surface \cite{Nandi2018,EslamPanah2019a,Li2021}.
An important quantity representing the SQM phase is the ratio of energy density to the baryon number density, $\epsilon / n_B$ at zero pressure. The value of $\epsilon / n_B$ should be less than that of the most stable nuclide ($^{56}Fe$), $\epsilon / n_B=930MeV$. While this condition is expected to occur at densities greater than $n_B=0.16fm^{-3}$, we show that the above condition ($\epsilon / n_B < 930 MeV$)  can occur at lower densities in the models used in this paper. This happens by employing a set of equations coming from the chemical equilibrium and the charge neutrality conditions and also imposing the following constraints for SQM.
I) The ratio of $\epsilon / n_B$ at zero pressure should be lower than $930MeV$.
II) The perturbative term has to be lower than the free term.
III) The number density of strange quark must be non-zero.
The onset density is obtained to be $n_B\sim0.1fm^{-3}$ which shows that SQM phase can occur in $n_B<0.16fm^{-3}$.
in Ref. \cite{Kurkela2010}, by using perturbative calculation, the same behavior is shown to be true ($n_B\sim0.15fm^{-3}$). The reason for obtaining a lower density in this paper is due to the fact that we use a  modified  QCD running coupling constant. However, one might say that the onset density of the quark matter might be greater than the value obtained from the leading order perturbative calculations, and considering the higher-order terms in perturbation leads to higher values for onset density of SQM. For this issue, we have also used other values of onset density of SQM, including 0.128 $fm^{-3}$ and $0.155fm^{-3}$.
We do not consider confining effects for the following reason. The static potential $V(r)$ is divided into two parts, including
$V_{NP}(r)$ and $V_{GE}(r)$ as the non-perturbative (NP) confining potential  and the gluon-exchange term, respectively.  $V_{NP}(r)$ appears at $q\bar{q}$ separations for $r > T_g$, where $T_g$ is the gluonic correlation length, $T_g\sim0.2fm^{-3}$ \cite{Badalian2005}. At zero temperature and finite chemical potential with $N_{f}$ flavor, the phenomenological models suggest $Q=[1\overline{\mu}-4\overline{\mu}] (\overline{\mu}\equiv\sum_{f}\mu _{f}/3)$ \cite{Schneider2003,Karmakar2019,Bandyopadhyay2019,Kurkela2010}. We set the maximum value for $Q$ ($Q=4\overline{\mu}$) to find the maximum possible mass of the SQS in our models and minimize the confinement effects. However, we found that for $Q\gtrsim 3.4(\mu_u+\mu_d+\mu_s)/3$, the results for the structural properties of SQS do not change considerably (see appendix C for more details). By considering $Q=4(\mu_u+\mu_d+\mu_s)/3$, the onset densities $0.1 fm^{-3}$, $0.128 fm^{-3}$ and $0.155 fm^{-3}$ lead to the range of $r\lesssim0.206fm$, $r\lesssim0.195fm$ and $r\lesssim0.185fm$, respectively. The value of $r$ is speculated by the uncertainty relation ($r\times Q\simeq\hbar$).  By solving charge neutrality and beta equilibrium equations for each onset density, we obtain the minimum chemical potential values for each quark flavor and, consequently, the minimum value of $Q$. Then by uncertainty relation, the maximum value of $r$ is obtained. It is worth mentioning that in Ref. \cite{Kurkela2010}, it is discussed that for the renormalization scales, $Q<0.8GeV$, the perturbative calculations are unreliable due to different uncertainties in the values of strange quark mass and coupling constant. Meanwhile in our calculations, the onset densities $0.1fm^{-3}$, $0.128 fm^{-3}$ and $0.155 fm^{-3}$ correspond to \textcolor{black}{$Q>0.955GeV$, $Q>1.01GeV$ and $Q>1.06GeV$}, which shows that our results are reliable.
\section{{IR behavior of $\protect\alpha _{s}$ in different models}}
\begin{figure}[tbph]
\centering
\par
\includegraphics[width=8cm,height=6cm]{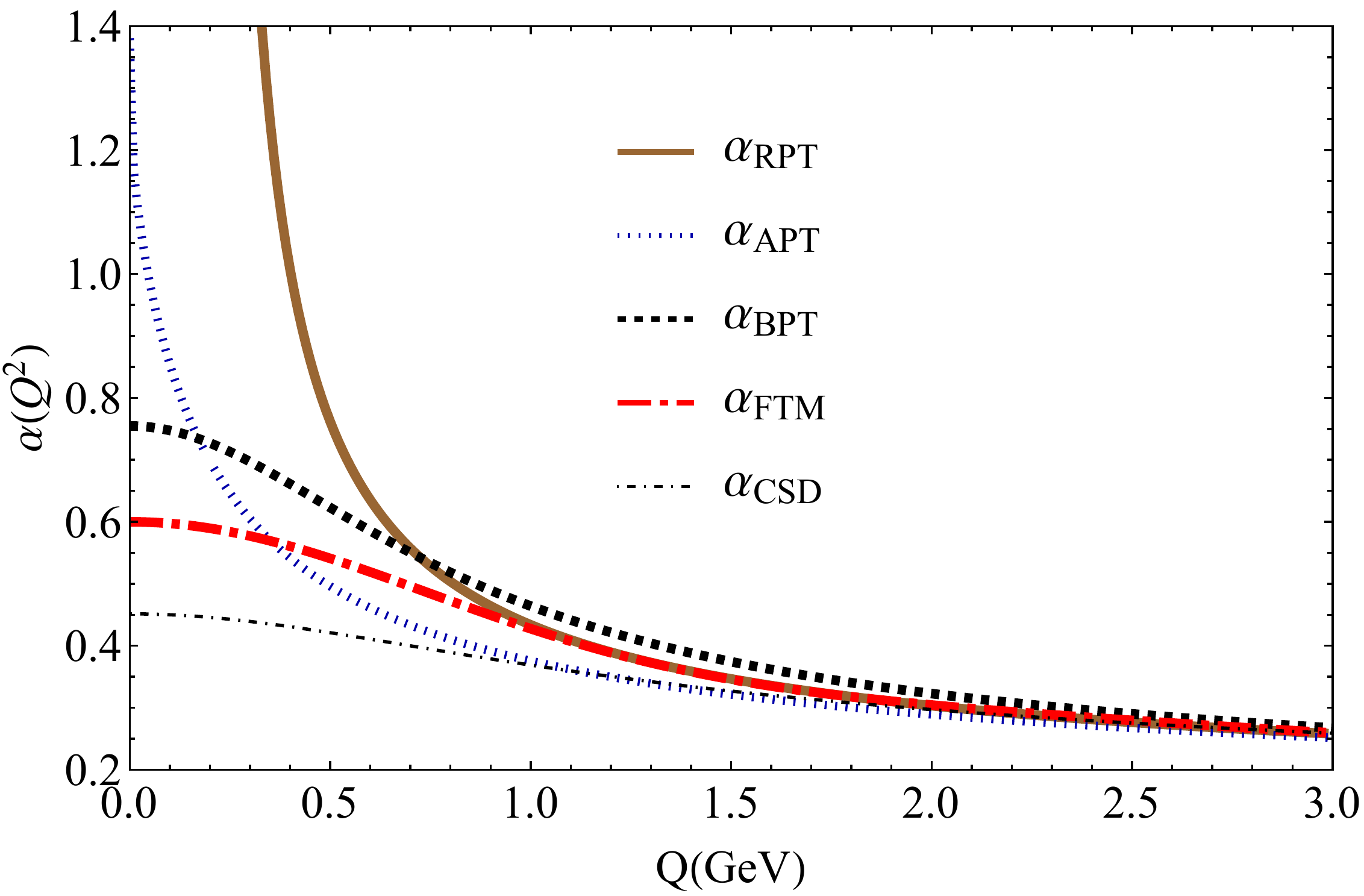}
\caption{QCD running coupling constant versus renormalization scale for
different models.}
\label{coupling diagram}
\end{figure}

As discussed above, there are experimental evidences and models which
support the slowly varying or freezing behavior of the running coupling
constant in the IR region. However, there is no consensus on the freezing
point and the corresponding value of the coupling constant. In this section,
the running coupling constant behaviors of different QCD models in the IR $%
Q^{2}$ are investigated. These models are as follows: i) regular
perturbation theory (RPT); ii) APT; iii) BPT; iv) flux-tube model (FTM) \cite%
{Godfrey1985}; and v) Cornwall Schwinger-Dyson equation (CSD) \cite%
{Cornwall1982}. The behavior of the running coupling constant in these
models for $Q<3GeV$ is shown in Figure. \ref{coupling diagram}. According to
this figure, as $Q$ decreases in the IR region, $\alpha _{APT}$ varies much
more slowly than $\alpha _{RPT}$. Nevertheless, compared to the other
models, $\alpha _{APT}$ increases faster in the IR region, especially for $%
Q<0.7$$GeV$. The couplings of the BPT, FTM, and CSD models have so slowly
varying behavior in the IR momenta that it can be said that they freeze
compared to the couplings of RPT and APT. However, for ultraviolet (UV)
momenta, all the models have an asymptotic behavior and coincide with each
other. Two of the mentioned coupling constants (except for that of RPT) with
higher values in the IR region are used. After a brief description of these
models, they are used to calculate the perturbative EOSs of SQM. The models
are as follows:

$1$) APT: The analytic coupling constant ($\alpha _{an}$) at one-loop
approximation \cite{Shirkov1997} is derived in Appendix A as
\begin{equation}
\alpha _{APT}^{(1)}\left( Q^{2}\right) =4\pi /\beta _{0}\left[ \left( \ln
\left( \frac{Q^{2}}{\Lambda ^{2}}\right) \right) ^{-1}+\frac{\Lambda ^{2}}{%
\Lambda ^{2}-Q^{2}}\right] .
\end{equation}

Supposing that $N_{c}$ and $N_{f}$ are the number of colors and flavors
respectively, $\beta _{0}$ is defined as $\frac{11N_{c}-2N_{f}}{3}$. It can
be seen that this running coupling constant has $Q^{2}$ analyticity at all
points and decreases monotonously in the IR region without any divergence
\cite{Deur2016}. The interesting feature of this running coupling is that $%
\alpha _{an}\left( Q^{2}=0\right) =4\pi /\beta _{0}$ has no dependence on
the QCD scale parameter ($\Lambda $). Furthermore, all perturbative orders
of $\alpha _{an}$ have the same value at $Q^{2}=0$. Hence, it is expected
that different orders of $\alpha _{an}\left( Q^{2}\right) $ have close
values. For example, in the $\overline{MS}$ modified minimal subtraction
scheme, $\alpha _{an}^{(1)}\left( Q^{2}\right) $ and $\alpha
_{an}^{(2)}\left( Q^{2}\right) $ differ within the $10\%$ interval and $%
\alpha _{an}^{(2)}\left( Q^{2}\right) $ and $\alpha _{an}^{(3)}\left(
Q^{2}\right) $ differ within the $1\%$ interval \cite{Shirkov1997}.

$2$) BPT: The running coupling constant in the framework of the BPT is given
by
\begin{equation}
\alpha _{BPT}^{(1)}(Q^{2})=\frac{4\pi }{\beta _{0}}\left( \ln \left[
(Q^{2}+m_{2g}^{2})/\Lambda ^{2}\right] \right) ^{-1},
\end{equation}
where $m_{2g}^{2}$ is the mass of two gluons connected by the fundamental
string ($\sigma $). This effective mass is added to the logarithm argument
to avoid the $\Lambda $ pole problem (see Appendix B for more details). In
the $q\overline{q}$ potential \cite{Simonov1995}, $m_{2g}^{2}=2\pi \sigma
\approxeq 1GeV$ \cite{Simonov2011,Badalian2019}.

\subsection{Thermodynamic potential}
The system under study is the SQM composed of up, down, and strange quark
flavors at zero temperature and finite chemical potential. The masses of the
up and down quarks are negligible. However, for the strange quark, the
running mass with the value of $0.95_{-3}^{+9}MeV$ $\cong 0.1GeV$ in $Q=2GeV$
\cite{Zyla2020} is considered (see Eq. \ref{running mass}). Since the QED interactions between quarks and
the interactions between gravitons are negligible compared to the QCD
interaction, QCD is the dominant interaction in the system. By knowing the
thermodynamic potential of the system, the properties of the SQM, such as
energy density, quark number density, pressure, sound speed, adiabatic
index, etc., can be derived. To obtain the thermodynamic potential ($\Omega $%
), it is divided into non-interacting and perturbative parts: i) the
non-interacting or free part consists of non-interacting quarks and
electrons; ii) the perturbative part consists of the QCD interactions
between the quarks. The perturbative part has been previously obtained
through two- and three-loop Feynman diagrams up to the first and second
order of the coupling constant, in Refs. \cite{Fraga2006} and \cite%
{Kurkela2010}, respectively. \textcolor{black}{The thermodynamic potential of a quark flavor with mass $m$ up to leading order is as follows \cite{Kurkela2010}
\begin{equation}
-\frac{\Omega}{V}=\sum_{{N_f=1}}^{3}\left({\mathcal M}_1+\frac{{\mathcal M}_2\alpha_s(Q)}{4\pi}\right),
\end{equation}
where ${\mathcal M}_1$ and $\frac{{\mathcal M}_2\alpha_s(Q)}{4\pi}$ are  non-interacting and perturbative parts, respectively. ${\mathcal M}_1$ and ${\mathcal M}_2$ are as follows
\begin{equation}
{\mathcal M}_1=
\frac{N_c \mu^4}{24\pi^2}\bigg\{2\hat{u}^3-3z \hat{m}^2\bigg\},\label{m1}
\end{equation}
\begin{equation}
{\mathcal M}_2= \frac{d_A \mu^4}{4\pi^2}\Bigg\{-6z \hat{m}^2 \ln\frac{Q}{m} +2\hat{u}^4 - 4z \hat{m}^2 -3z^2 \Bigg\}, \label{m2}
\end{equation}
where, $\hat{u}\equiv(\sqrt{\mu^2-m^2})/\mu$, $\hat{m}\equiv m/\mu$, $z \equiv \hat{u}-\hat{m}^2\,\ln\bigg[\frac{1+\hat{u}}{\hat{m}}\bigg]$ and $d_A\equiv N^2_c-1$.
As we mentioned, we consider up and down quarks massless. The running mass of the strange quark is \cite{Kurkela2010}
\begin{equation}
m(Q)=m(2GeV)\left(\frac{\alpha_s(Q)}{\alpha_s(2GeV)}\right)^{\gamma_0/\beta_0},\label{running mass}
\end{equation}
where $\gamma_0\equiv3\frac{N_c^2-1}{2N_c}$ and $\beta_0\equiv\frac{11N_c-2N_f}{3}$. Moreover, we have considered $m(2GeV)\simeq0.1GeV$ and $\alpha_s(2GeV)\simeq0.3$ \cite{Zyla2020}.}
The pressure is derived from the relation $P=-B-\frac{\Omega }{V}$, where $V$ is the volume of the system. It is notable
that $B$ is a free parameter considered for all non-perturbative effects not
included in the perturbative expansion \cite{Kurkela2010}. \textcolor{black}{The values of $B$ are obtained in such
a way to get zero total pressure at the surface of the star
where the baryon number density is minimum \cite{Kurkela2010,sedaghat2022plb}. Therefore the B parameter corresponds to the onset density (see tables \ref{MRI} and \ref{MRII}). If the
perturbative interaction between the quarks is neglected, $B$ will have the
role of the bag constant in the system.} The renormalization scale $Q$
appears in pressure through the running mass of the strange quark and the
running coupling constant. At finite temperature $(T)$ and finite chemical
potential $(\mu )$, the phenomenological models suggest $Q=2\pi \sqrt{T^{2}+%
\frac{\mu ^{2}}{\pi ^{2}}}$ for a massless quark. At zero temperature with $%
N_{f}$, $Q=2\left( {\ \sum_{f}\mu _{f}/3}\right) \equiv 2\overline{\mu }$
which can vary by a factor of $2$ with respect to its central value \cite%
{Schneider2003,Karmakar2019,Bandyopadhyay2019}. \textcolor{black}{Employing all the required constraints
for SQM (charge neutrality, beta equilibrium, $\epsilon/nB<0.930 MeV$, and $n_s>0$), we obtain the EOS for SQS.} \newline

\subsection{Stability conditions and EOSs}

After calculating the thermodynamic potential, the EOSs of the stable SQSs
can be obtained from the following relation
\begin{equation}
\varepsilon =-\left( P_{u}+P_{d}+P_{s}\right) +\mu _{u}n_{u}+\mu
_{d}n_{d}+\mu _{s}n_{s},  \label{EoS}
\end{equation}%
where $\varepsilon $, $P_{i}$, $\mu _{i}$ and $n_{i}$ are the energy
density, pressure, chemical potential, and quark number density with flavor $%
i$, respectively. It should be noted that to calculate the EOSs, the charge
neutrality and beta equilibrium constraints have to be imposed \cite
{Blaschke2001}. As a result, $\mu _{s}=\mu _{d}\equiv \mu ,~\mu _{u}=\mu
-\mu _{e},~$and$~2n_{u}/3-n_{d}/3-n_{s}/3-n_{e}=0$ hold where $n_{e}=eB\mu
_{e}^{2}/(2\pi )^{2}$is the electron number density. Another constraint is
that the minimum energy density per baryon number density should be lower
than that of the most stable nucleus ($^{56}Fe$) \cite%
{Witten1984,Terazawa1989,Weber2005}, i.e. $\varepsilon /n_{B}\leq 0.93GeV$
where $n_{B}$ is the baryon number density and is equals $n_{B}=\left(
n_{u}+n_{d}+n_{s}\right) /3$. The minimum value of $n_{B}$ corresponds to
the point that the strange quark number density is non-zero provided that
the constraint for the minimum energy density per baryon is satisfied and
the perturbative expansions of the pressures and the quark number densities
are not broken down.

\subsection{Thermodynamic properties of SQM in BPT and APT models}

Here, the results for the EOSs of SQM in the APT and BPT models are
presented. First, the critical baryon number density ($n_{cr}$) at which SQM
begins to appear should be obtained. For this purpose, the minimum value of $%
n_{B}$ is calculated. For this value, the perturbative expansion must be
valid, and the constraints of the SQM must be satisfied. The obtained baryon
number density is referred to as the minimum critical baryon number density
and is denoted by $(n_{cr})_{min}$. The value of $n_{cr}$ must be more than $%
(n_{cr})_{min}$ because the higher order contributions (compared with the
leading order) in the perturbative expansion are ignored. Table. \ref{min-nB}%
, shows the values of $(n_{cr})_{min}$ in the RPT, APT and BPT models for
different values of the renormalization scale. As one can see from Table. %
\ref{min-nB}, the $(n_{cr})_{min}$ of the APT and BPT models is obtained at
lower energies than that of the RPT model. This feature is due to the
behavior modification of the coupling constant in the IR region in the APT
and BPT models. Moreover, there is little difference between the $%
(n_{cr})_{min}$ values of the APT, and BPT models. This stems from the fact
that the $(n_{cr})_{min}$ values are obtained at $Q\gtrsim 0.8GeV$ where the
coupling constants in the APT and BPT models behave almost similarly (see
figure. \ref{coupling diagram}). According to Table. \ref{min-nB}, the value
of $(n_{cr})_{min}$ for the APT and BPT models is about $0.1fm^{-3}$.
As mentioned above, $n_{cr}>(n_{cr})_{min}$ values must be used. Using bag
models in Refs. \cite{Sagert2009} and \cite{Sagert2010}, $n_{cr}$ was
obtained in the interval $0.1fm^{-3}\lesssim n_{cr}\lesssim 0.5fm^{-3}$. By
setting $Q=4\overline{\mu}$, different values of $n_{cr}\geq
(n_{cr})_{min}$ are selected to obtain EOSs, and consequently, the maximum
gravitational masses of SQSs in the APT and BPT models. 
It is noteworthy that the maximum gravitational masses are decreased by increasing $n_{cr}$.

\begin{table}[h]
\caption{\textcolor{black}{Minimum values of baryon number density $(n_{cr})_{min}$ for the
strange quark matter by considering $Q=4\overline{\mu}$ in various models.}}
\label{min-nB}\centering
\par
\begin{tabular}{|c|cc||cc||cc||}
\hline
& $RPT$ & $RPT$ & $APT$ & $APT$ & $BPT$ & $BPT$ \\ \hline
& $\overline{\mu}$ & $(n_{cr})_{min}$ & $%
\overline{\mu}$ & $(n_{cr})_{min}$ & $\overline{\mu}$ & $(n_{cr})_{min}$ \\
& (MeV) & ($fm^{-3}$) & (MeV) & ($fm^{-3}$) & (MeV) & ($fm^{-3}$) \\ \hline
& \textcolor{black}{278} & \textcolor{black}{0.154} & \textcolor{black}{236} & \textcolor{black}{0.099}& \textcolor{black}{245}&\textcolor{black}{0.097} \\ \hline
\end{tabular}
\end{table}

\begin{figure}[h]
\center{\includegraphics[width=8.5cm] {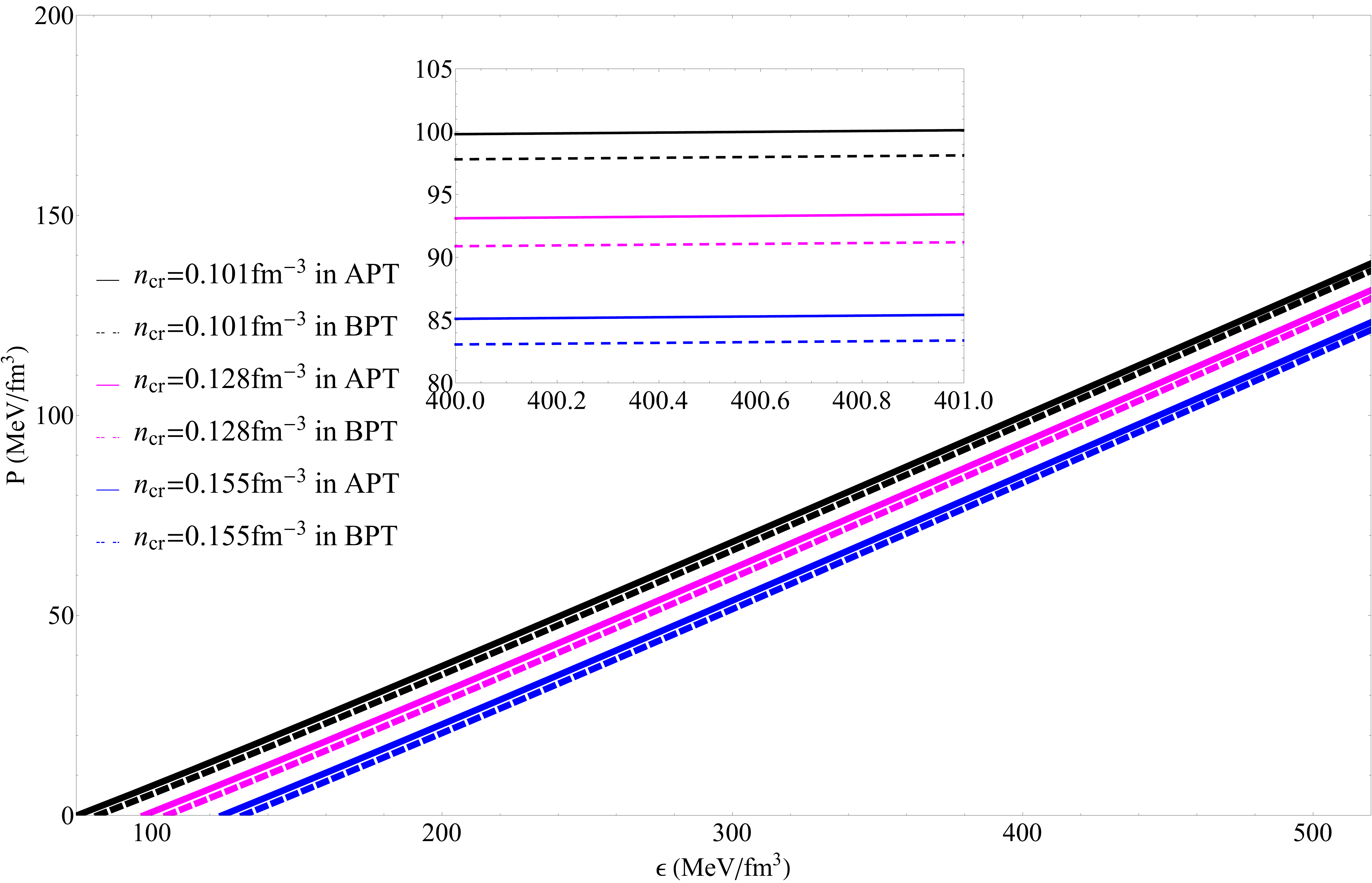}}
\caption{\textcolor{black}{EOSs of SQM in APT and BPT models for different choices of $n_{cr}$%
. Continuous and dashed lines correspond to APT and BPT models,
respectively. Each color indicates a $n_{cr}$ as follows. Continuous and
dashed black lines correspond to $n_{cr}=0.101fm^{-3}$, continuous and
dashed magenta lines correspond to $n_{cr}= 0.128fm^{-3}$, and continuous
and dashed blue lines correspond to $n_{cr}= 0.155fm^{-3}$.}}
\label{eos}
\end{figure}

In Figure. \ref{eos}, the EOSs of SQM in the APT and BPT models are
presented for different values of $n_{cr}$. Each color corresponds to a
different $n_{cr}$. In both models, the EOSs become softer when $n_{cr}$
increases.

\begin{figure}[h]
\center{\includegraphics[width=8.5cm] {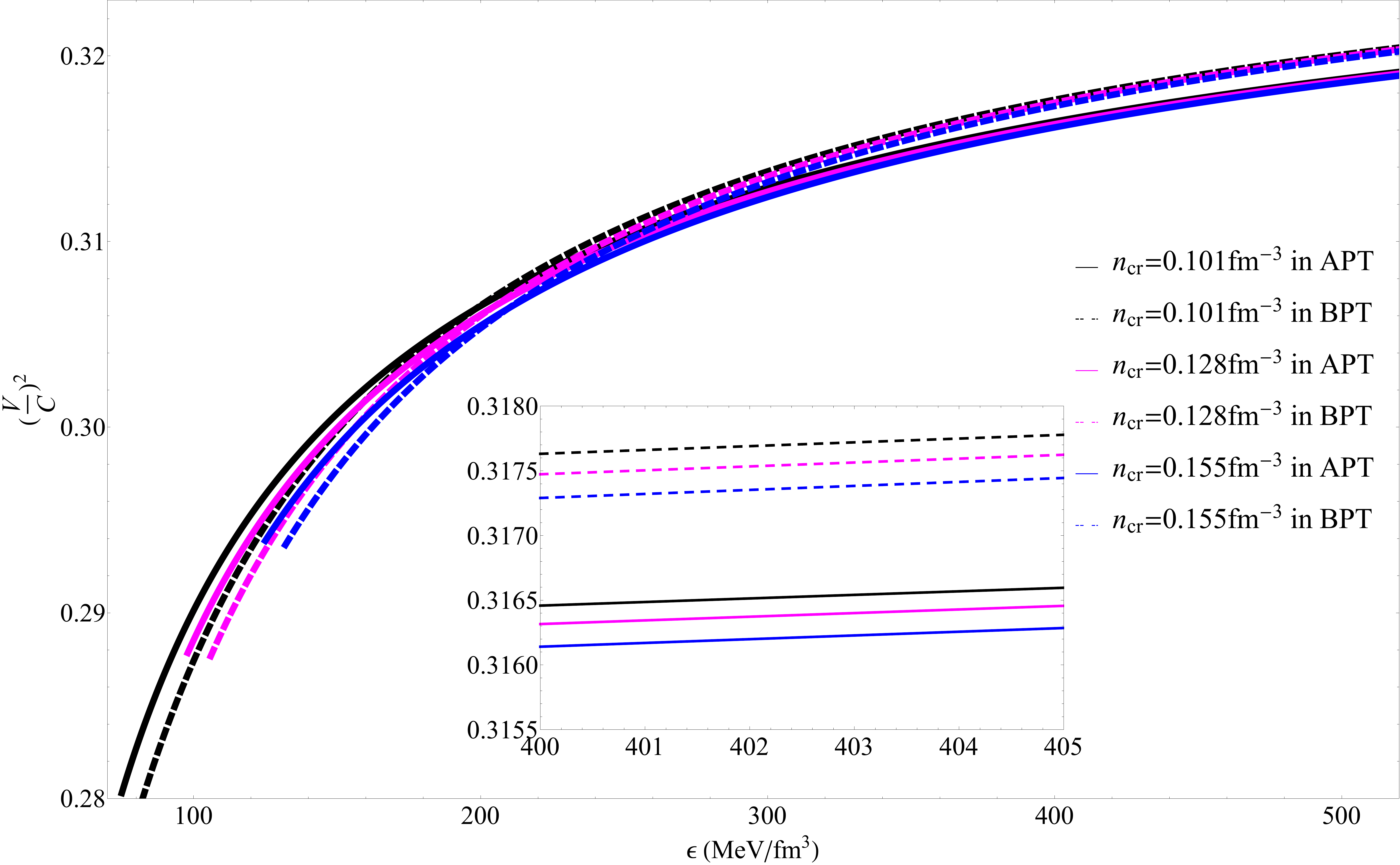}}
\caption{\textcolor{black}{Squared sound velocity versus energy density in APT and BPT models
for different choices of $n_{cr}$. Continuous and dashed lines correspond to
APT and BPT models, respectively. Each color indicates a $n_{cr}$ as
follows. Continuous and dashed black lines correspond to $%
n_{cr}=0.101fm^{-3} $, continuous and dashed magenta lines correspond to $%
n_{cr}= 0.128fm^{-3}$, and continuous and dashed blue lines correspond to $%
n_{cr}=0.155fm^{-3}$.}}
\label{sound}
\end{figure}

It is worthwhile to mention that the EOSs should satisfy the conditions of
causality and dynamic stability. To meet the causality condition, the speed
of sound ($c_{s}=\sqrt{dP/d\epsilon }$) should be lower than the speed of
light in vacuum ($c_{s}\leq 1$). Figure. \ref{sound}, shows the behavior of $%
c_{s}^{2}$ versus energy density for different values of $n_{cr}$. Based on
some arguments in the perturbation theory, $c_{s}^{2}$ should approach $1/3$
from below at asymptotically large densities \cite{Tan2020}. Figure. \ref%
{sound}, indicates that this behavior is well satisfied in our EOSs for both
APT and BPT models.

To satisfy the dynamic stability condition, the value of the adiabatic index
$\left( \Gamma =\frac{(P+\epsilon )dP}{Pd\epsilon }\right) $ should be
higher than $4/3$ \cite%
{Chandrasekhar1964,Bardeen1966,Kuntsem1988,Mak2013,Eslam2017}%
. This constraint holds for our EOSs in both APT and BPT models. Figure. \ref%
{adia}, shows that the adiabatic index of SQM versus energy density is
higher than 4/3 for all $n_{cr}$ values used in both APT and BPT models. It
is worth mentioning that the constraints $\epsilon +P\geq 0$ and $\epsilon
\geq \left\vert P\right\vert $ \cite%
{Hendi2016,Hendi2017,EslamPanah2019b,Roupas2021} are well satisfied in our
calculations as well.

\begin{figure}[h]
\center{\includegraphics[width=8.5cm] {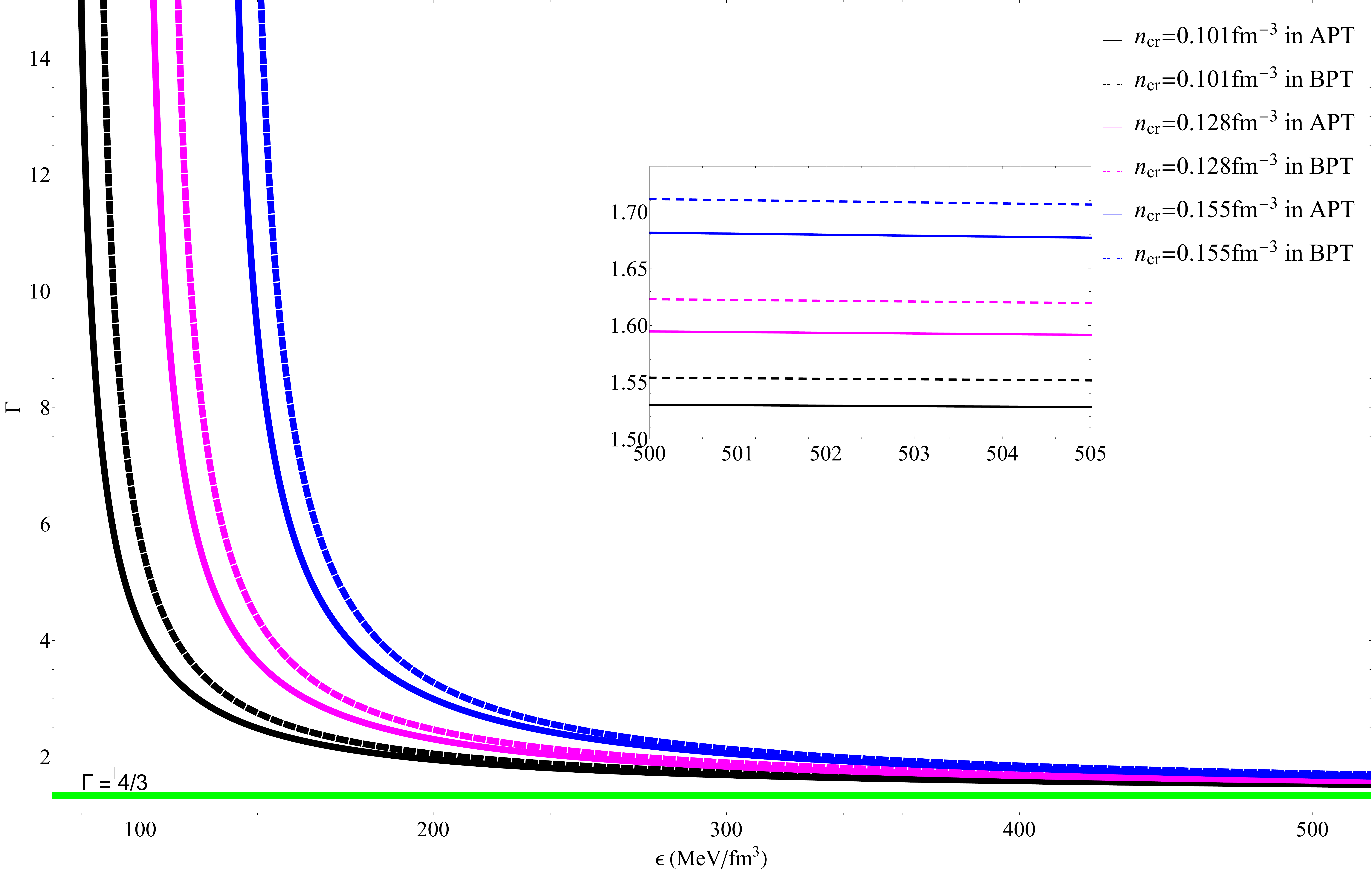}}
\caption{\textcolor{black}{Adiabatic index in APT and BPT models for different choices of $%
n_{cr}$. Continuous and dashed lines correspond to APT and BPT models,
respectively. Each color indicates a $n_{cr}$ as follows. Continuous and
dashed black lines correspond to $n_{cr}=0.101fm^{-3}$, continuous and
dashed magenta lines correspond to $n_{cr}=0.128fm^{-3}$, and continuous and
dashed blue lines correspond to $n_{cr}=0.155fm^{-3}$.}}
\label{adia}
\end{figure}

\subsection{Calculation of the remnant masses of non-rotating and
rapidly-rotating compact stars}

The gravitational wave (GW) observations give the total gravitational mass
of the binary system with the assumption that its components are at an
infinite distance from each other \cite{Gao2020}. In order to obtain the
remnant mass of a binary NS merger, the total baryonic mass must first be
calculated from the total gravitational mass through the conversion
relations shown in Table. \ref{universal-relations} \cite%
{Lattimer1989,Timmes1996,Coughlin2017,Gao2020}. As one can see from Table. %
\ref{universal-relations}, there are some conversion relations between
baryonic mass ($M_{b}$) and gravitational mass ($M_{g}$) for NR and rapidly
rotating (RR) NSs. After calculating the total baryonic mass from components
of the binary, the mass ejected during the merger is deduced from it. Hence,
the remnant baryonic mass is obtained and converted back to the remnant
gravitational mass. Before the merger, the low-spin approximation is assumed
for the components in the binary system. Therefore, the conversion relations
are used for NR stars. There is a different scenario for the remnant mass.
Since the remnant mass must be rapidly spinning, the RR conversion relations
between $M_{b}$ and $M_{g}$ are used \cite{Gao2020}. The first and last
relations in Table. \ref{universal-relations} are used as the conversion
relations for the phases before and after the merger, respectively. Assuming
the maximum expected value of the mass ejected for GW190425 ($M_{ejecta}=0.04%
{M}_{\odot }$ \cite{Foley2020}), the remnant mass of GW190425 is obtained in
the range of $3.11-3.54{M}_{\odot }$. It should be noted that in the
calculations, the mass ratio of the binary GW190425 is bound to $0.4-1$ \cite%
{Abbott2020ApJL}. \textcolor{black}{The remnant might require the EOS at non-zero temperature and out of beta equilibrium \cite{Khadkikar2021}, and consequently, the universal relations at finite temperature are needed \cite{N.khosravi2021}. But it depends on the star's state and the system's Fermi temperature. For the strong shocks at the merger, the temperature of the merged stellar object can reach up to $30-50MeV$ \cite{L.Baiotti2017}.
However, the star's temperature will be rapidly reduced by neutrino emission \cite{L.Baiotti2017}. On the other hand, the Fermi temperature of quark stars is several times the Fermi temperature of neutron stars. As we know, the Fermi energy at zero temperature is equivalent to the chemical potential. Table \ref{min-nB} shows that the lowest value for the Fermi energy is about $240MeV$. While in a neutron star at saturation density, this value is about 50$MeV$. Therefore, the finite temperature effects can be considerable for the newborn neutron stars \cite{Khadkikar2021}.\\}
\begin{table}[h]
\caption{Relation between total baryonic mass and total gravitational mass
for NR and RR stars. Below each relation, there are two numbers as the
maximum residual error (outside the parenthesis) and the average residual
error (inside the parenthesis).}
\label{universal-relations}\centering
\par
\begin{tabular}{||c|c||}
\hline
NR NSs & RR NSs \\ \hline
$M_b=Mg+0.075M_g^2$ & $M_b=Mg+0.064M_g^2$ \\
$5.8\%(1.5\%)$ & $3.3\%(1.2\%)$ \\
\cite{Timmes1996} & \cite{Gao2020} \\ \hline
$M_b=Mg+0.084M_g^2$ & $M_b=Mg+0.073M_g^2$ \\
$4.1\%(1.7\%)$ & $6.0\%(1.6\%)$ \\
\cite{Lattimer1989} & \cite{Gao2020} \\ \hline
$\frac{M_b}{M_g}=1+0.89(\frac{M_g}{R})^{1.2}$ & $M_b=Mg+.056M_g^2+0.002M_g^3$
\\
$2.6\%(0.56\%)$ & $4.0\%(0.95\%)$ \\
\cite{Coughlin2017} & \cite{Gao2020} \\ \hline
\end{tabular}
\end{table}

\section{SQS structure}
\begin{figure}[h]
	\center{\includegraphics[width=8.5cm] {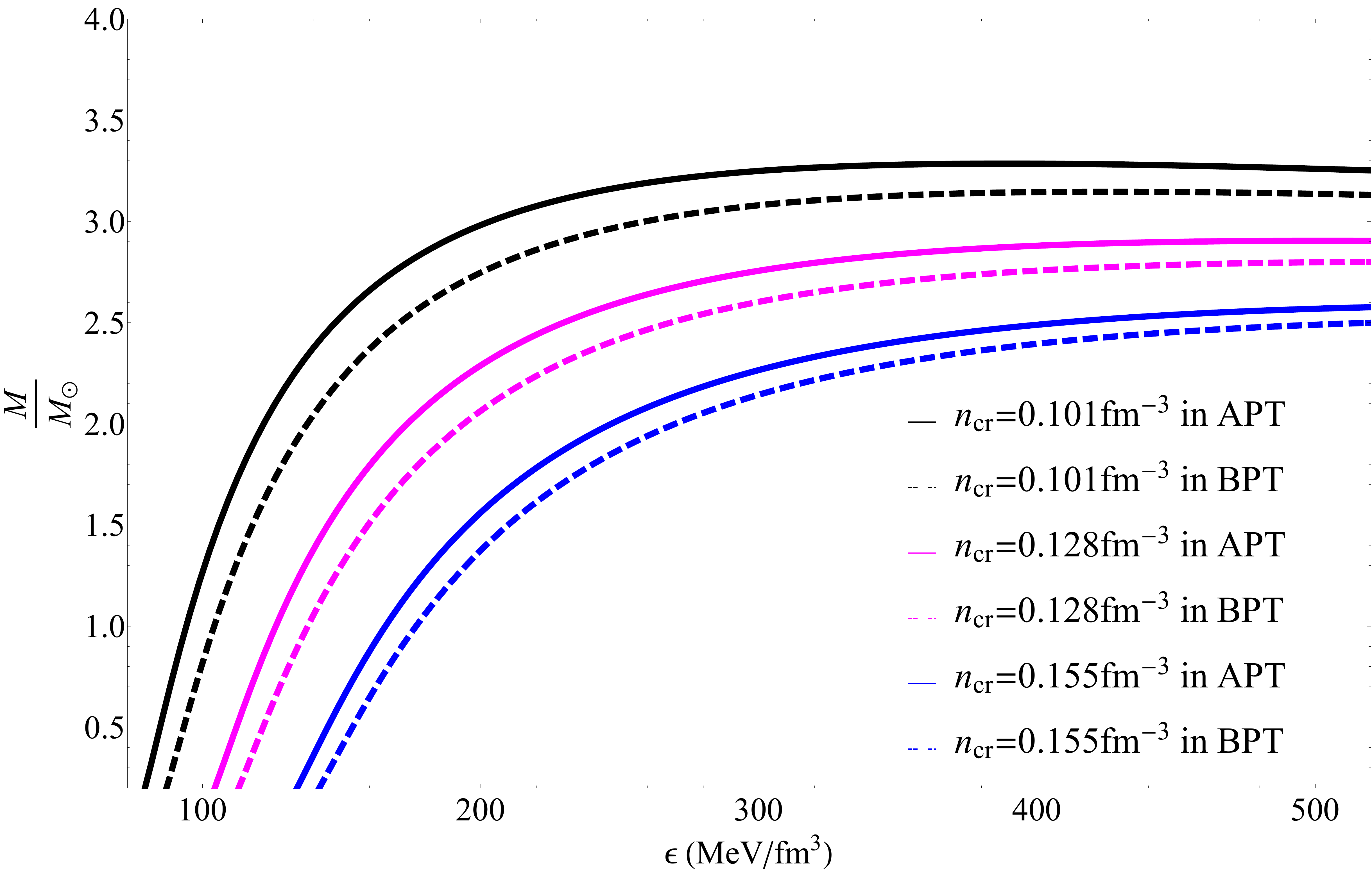}}
	\caption{\textcolor{black}{Mass versus energy density in APT and BPT models for different choices of $n_{cr}$. Continuous and dashed lines correspond to APT and BPT models,
		respectively. Each color indicates a $n_{cr}$ as follows. Continuous and
		dashed black lines correspond to $n_{cr}=0.101fm^{-3}$, continuous and
		dashed magenta lines correspond to $n_{cr}=0.128fm^{-3}$, and continuous and
		dashed blue lines correspond to $n_{cr}=0.155fm^{-3}$.}}
	\label{mass-energy}
\end{figure}

\begin{figure*}[tbp]
\center{\includegraphics[width=12cm]
{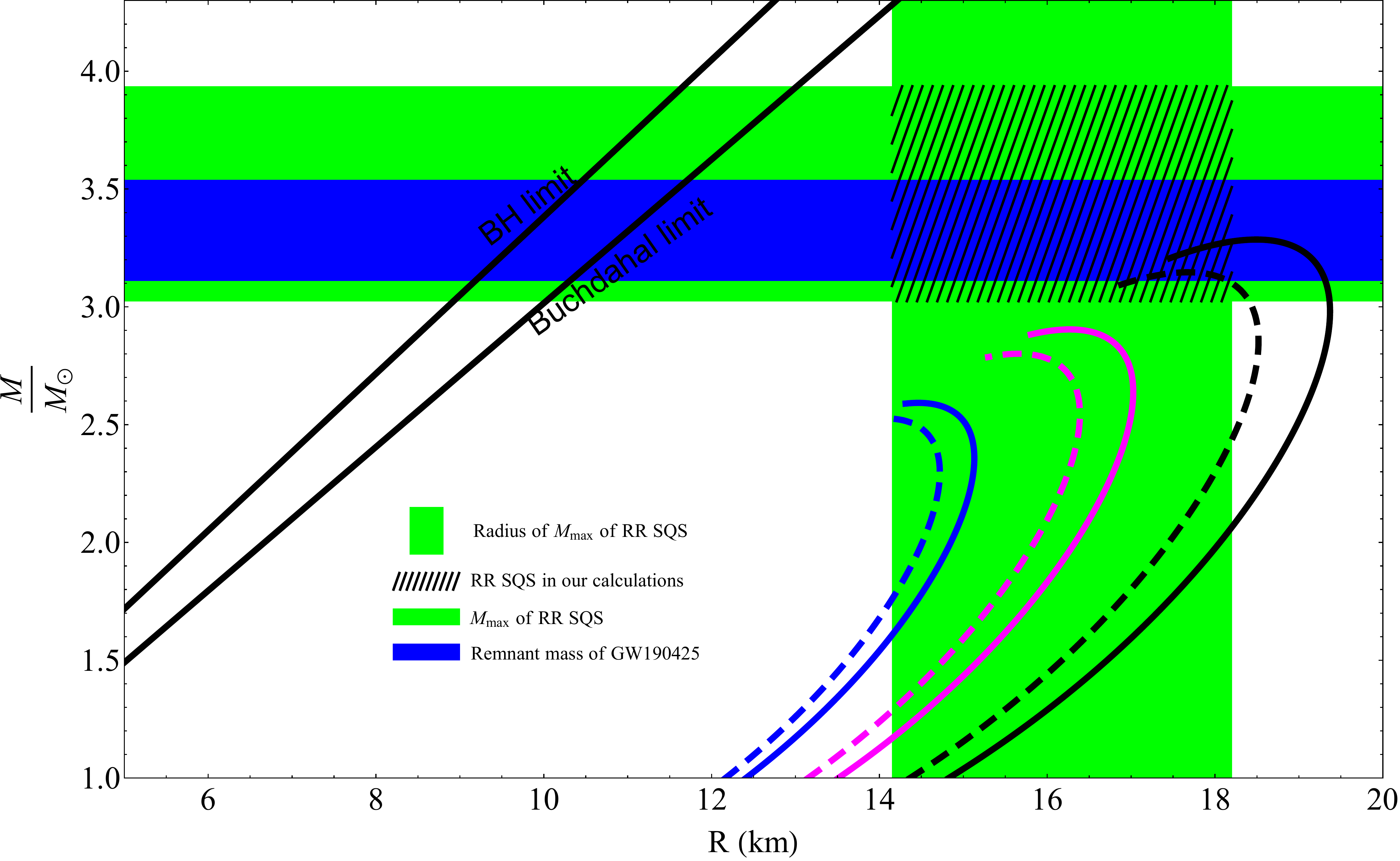}}
\caption{\textcolor{black}{$M$-$R$ of SQSs in the APT and BPT models for different choices of $%
n_{cr}$. The blue region is related to the remnant mass of GW190425. The
horizontal and vertical green regions show intervals of maximum mass of RR
SQSs and corresponding radii, respectively. The black hatched region is
related to our results for maximum masses and corresponding radii of RR
SQSs. Continuous and dashed lines correspond to different $n_{cr}$, which
are the same as the ones in Figures. \protect\ref{eos}, \protect\ref{sound}
and \protect\ref{adia}.}}
\label{m-r}
\end{figure*}

Using the computed EOSs in section C, the maximum gravitational mass can be obtained by solving the TOV equation \cite{Tolman1939,Oppenheimer1939}. \textcolor{black}{The results for the APT and BPT models have been presented in Tables. \ref{MRI} and \ref{MRII}, and also figures. \ref{mass-energy} and \ref{m-r}. The limiting behavior of the mass in Figure \ref{mass-energy} shows the maximum gravitational mass of SQS. From this figure, we can obtain the range of energy density for SQS.} In Ref. \cite{Kurkela2010}, the maximum gravitational mass of SQS has been obtained as $M\simeq 2.75{M}_{\odot}$ by using the RPT model. Our results in the APT model for $(n_{cr})_{min}=0.101fm^{-3}$ and $n_{cr}=0.128fm^{-3}$ are considerably larger than that in \cite{Kurkela2010}. This indicates the great importance of the IR behavior of the running coupling constant in the structure of a quark star.
\textcolor{black}{It is notable that the maximum gravitational masses obtained in our models are outside the limits obtained by observational constraints for neutron stars (such as GW170817 \cite{Rezzolla2018,Shibata2019}).  Such a violation  motivates us to study the remnant mass of GW190425 as a strange quark star. However, the two-families scenario \cite{Alessandro Drago} (neutron stars and strange quark stars coexist) allows the existence of massive quark stars without refuting observational arguments. Furthermore, the compact object with the mass $2.5 - 2.67 M_{\odot}$ in GW190814 is expected to be a strange quark star \cite{Zhiqiang Miao}. In Ref. \cite{I.Bombaci2021}, this object is investigated as a SQS within the two-families scenario.} There are different colored regions in Figure. \ref{m-r}. The blue region shows the remnant mass of GW190425 obtained in section D. In the current study, the maximum gravitational mass for the RR SQS represented by the horizontal green region ranged from \textcolor{black}{$3.02 - 3.94$${M}_{\odot }$}. The vertical green region, which represents the corresponding radius of the maximum mass of the RR SQSs ranges from \textcolor{black}{14.14$km$ to $18.52$$km$} in our
calculations. Finally, the black hatched region, the common region
between the two mentioned green regions, shows our results for the RR SQSs.
As can be observed in Fig. \ref{m-r}, this region completely covers the
remnant mass of GW190425.
\begin{table}[tbp]
\caption{The structural properties of the NR and RR SQSs in the APT model for different
values of $n_{cr}$. It is notable that indices NR and RR, are related to NR
and RR SQSs.}
\label{MRI}\centering
\par
\begin{tabular}{||c|c|c|c|c|c|c|c||}
\hline
$n_{cr}$ & $M_{TOV}$ & $R_{NR}$ & $M_{max_{RR}}$ & $M_{B_{NR}}$ & $%
R_{Sch_{NR}}$ & $z_{NR}$ & $\textcolor{black}{B}$ \\
$fm^{-3}$ & $({M}_\odot)$ & $(km)$ & $({M}_\odot)$ & $({M}_\odot)$ & $(km)$
&  & $\textcolor{black}{\frac{MeV}{(fm)^3}}$\\ \hline
0.101 & \textcolor{black}{3.28} & \textcolor{black}{18.52} & \textcolor{black}{3.94} & \textcolor{black}{5.58} & \textcolor{black}{9.67} & \textcolor{black}{0.45} & \textcolor{black}{15.23} \\ \hline
0.128 & \textcolor{black}{2.90} & \textcolor{black}{16.80} & \textcolor{black}{3.48} & \textcolor{black}{5.06} & \textcolor{black}{8.55} &\textcolor{black}{0.43}  & \textcolor{black}{20.31} \\ \hline
0.155 & \textcolor{black}{2.59} & \textcolor{black}{14.52} & \textcolor{black}{3.11} & \textcolor{black}{4.37} & \textcolor{black}{7.64} & \textcolor{black}{0.45} & \textcolor{black}{26.39} \\ \hline
\end{tabular}%
\end{table}
\begin{table}[tbp]
\caption{The structural properties of the NR and RR SQSs in the BPT model for different
values of $n_{cr}$.}
\label{MRII}\centering
\par
\begin{tabular}{||c|c|c|c|c|c|c|c||}
\hline
$n_{cr}$ & $M_{TOV}$ & $R_{NR}$ & $M_{max_{RR}}$ & $M_{B_{NR}}$ & $%
R_{Sch_{NR}}$ & $z_{NR}$ & $\textcolor{black}{B}$ \\
$fm^{-3}$ & $({M}_\odot)$ & $(km)$ & $({M}_\odot)$ & $({M}_\odot)$ & $(km)$
&  & $\textcolor{black}{\frac{MeV}{(fm)^3}}$\\ \hline
0.101 & \textcolor{black}{3.15} &\textcolor{black}{17.62}  &\textcolor{black}{3.78}  & \textcolor{black}{5.31} & \textcolor{black}{9.29} & \textcolor{black}{0.45} & \textcolor{black}{19.84} \\ \hline
0.128 &\textcolor{black}{2.80}  &\textcolor{black}{15.61}  &\textcolor{black}{3.36}  & \textcolor{black}{4.70} & \textcolor{black}{8.26} & \textcolor{black}{0.46} & \textcolor{black}{25.09}  \\ \hline
0.155 &\textcolor{black}{2.52}  &\textcolor{black}{14.14}  &\textcolor{black}{3.02}  & \textcolor{black}{4.26} & \textcolor{black}{7.43} & \textcolor{black}{0.45} & \textcolor{black}{31.03} \\ \hline
\end{tabular}
\end{table}

Now, it is explained how to calculate the maximum gravitational masses of
the RR SQSs in the APT and BPT models. As was mentioned in section D, the
remnant mass must be rapidly-rotating. If the rapidly-rotating remnant mass
loses its centrifugal support, it will collapse into a BH \cite{Sarin2020}.
\textcolor{black}{To obtain the maximum mass of the rotating SQSs, the universal relation
derived in Ref. \cite{Breu2016} was used. They used $15$
nuclear-physics EOSs for NR and RR compact stars and obtained a
universal relation between the maximum mass of a uniformly rotating
star ($M_{max_{RR}}$) and the maximum mass of the NR star obtained by TOV equation $(M_{TOV})$.
\begin{equation}
M_{max_{RR}}=(1.203\pm 0.022)M_{TOV}.
\end{equation}
As we mentioned, this universal relation has been obtained by nucleonic EoSs. It is notable that the authors of Ref. \cite{Breu2016} have also suggested the universal relation $M_{max_{RR}}=1.44 M_{TOV}$ for SQSs. But they have noted that additional work is needed to confirm this behavior. Therefore, we have not used this relation in our paper. However, if we use this universal relation, our results would cover better the remnant mass of GW190425.}

The result is a horizontal green region in Figure. \ref{m-r}, which ranges
from \textcolor{black}{$3.02 - 3.94$${M}_{\odot }$}. \textcolor{black}{It is noteworthy that considering anisotropic models for the quark star can even increase the mass up to $5M_{\odot}$ \cite{J.Horvath}.}

It has previously been shown that when the causality and Le Chatelier
principles are used in GR, the mass of a NS cannot be larger \ than $%
3.2M_{\odot }$ \cite{Rhoades1974}. Therefore, the remnant mass of GW190425
is not expected to be a NS. Moreover, the upper mass limit of a static
spherical star with a uniform density in GR is given by $M_{B}=\frac{4c^{2}R%
}{9G}$ (the Buchdahl theorem) \cite{Buchdahl1959}. According to this
compactness limit, a boundary between a BH and a star can be determined.
Indeed, for $M_{max}<M_{B}$, the compact object cannot be a BH. In other
words, the results show that the maximum masses of the compact objects in
the current study are less than the mass of the Buchdahl limit. Hence, these
massive compact objects are not BHs (see Tables. \ref{MRI} and \ref{MRII}).
In addition, the radii of these compact objects are more than the
Schwarzschild radius $\left( R>R_{Sch}=\frac{2GM}{c^{2}}\right) $ and the
obtained redshifts are less than $1$ $\left( z=\frac{1}{\sqrt{1-\frac{2GM}{%
c^{2}R}}}-1<1\right) $ (see Tables. \ref{MRI} and \ref{MRII}). These
results ensure that these compact objects are not BHs or NSs. As a result,
RR SQSs may have masses larger than $3.2M_{\odot}$. Therefore, the remnant
mass of GW190425 may be considered a SQS.

\section{Summary and Conclusion}

In this paper, the EOSs of SQSs were calculated in the leading order of $\alpha _{s}$ by using the APT and BPT
models. It is known that the constant coupling behavior of the QCD is a
challenging issue. Unlike the RPT model in which the coupling constant is
infinite at IR momenta, the employed models in this paper have finite values at all energy scales for the running coupling constant. Considering
a compact star as a large laboratory to probe the QCD models, the EOSs in GR were used in order to investigate the structural properties
of SQSs at zero temperature. We had not made any assumptions about the onset density of the quark matter. Rather, we obtained it by perturbative calculations. By employing a set of equations coming from the chemical equilibrium and the charge neutrality conditions and also imposing the constraints for SQM ($\epsilon / n_B < 930 MeV \ \& \ n_s>0$), we showed that the SQM phase could occur at $n_B<0.16fm^{-3}$ in leading order perturbation theory. We found the minimum value for onset density of SQM as $0.1fm^{-3}$. One should note that we study the matter inside a compact star or the matter after merging two neutron stars, which has completely different conditions than the usual conditions in laboratories. It is worth mentioning that in Ref\cite{Sagert2009}, the onset densities have been considered to be around $0.1fm^{-3}$ for stars with proton fractions $Y_p<0.3$. However, one might say that the onset density of the quark matter might be greater than the value obtained from the leading order perturbative calculations, and considering the higher-order terms in perturbation leads to higher values for onset density of SQM. For this issue, we have also used other values of onset density of SQM, including 0.128 $fm^{-3}$ and $0.155fm^{-3}$. Then, we obtained the maximum gravitational
masses of SQSs in APT and BPT models, which were considerably larger than that of the RPT model. By using the
component masses of  GW190425 as well as some conversion relations
between the baryonic mass and the gravitational mass, the remnant
mass of GW190425 was obtained. Our results for the maximum gravitational
mass of SQS were comparable with the remnant mass of GW190425. Then,
the obtained gravitational masses were modified by considering the effect
of the star's rotation on them. In this way, the results completely
covered the remnant mass of GW190425 and showed that the remnant mass of
GW190425 might be a SQS. Our results corresponded to three different onset densities including $0.101fm^{-3}$, $0.128fm^{-3}$ and $0.155fm^{-3}$. Even if we ignore our calculations corresponding to the onset density $0.101fm^{-3}$, our results still fall within the range of the remnant mass of GW190425 (we have obtained the remnant mass of GW190425 in the range $3.11-3.54M_{\odot}$). As tables 3 and 4 of the paper Show, for the onset density $0.128fm^{-3}$ \textcolor{black}{($Q>1.01GeV$), the maximum mass of SQS is \textcolor{black}{$3.48M_{\odot}$} and $3.36M_{\odot}$ in APT and BPT models, respectively.} Here, it can be asked the question of why
we compared our results with the remnant mass of GW190425. As the post-merger of GW190425 falls within the unknown mass gap region ($2.5 - 5M_{\odot}$), its nature
is ambiguous. Given the remnant mass of GW190425 $(3.11-3.54  M_{\odot })$,
it is unlikely to be a supermassive neutron star. Hence it may be a black hole
or a quark star. Our calculations suggest that the remnant mass of GW190425 may
be a strange quark star. In addition, if a pulsar falls within the range of
the obtained masses and radii, it is likely to be a strange quark star.

\section*{Acknowledgements}

We are indebted to Prof. Yu. A. Simonov and Prof. Luciano Rezzolla for their fruitful comments and discussions on the background perturbation theory and the maximum mass of rotating compact objects, respectively. SMZ and GHB thank the Research Council of Shiraz University. SMZ thanks the Physics Department of UCSD for their hospitality during his sabbatical. BEP thanks University of Mazandaran. The work of BEP has been supported by University of Mazandaran by title "Evolution of the masses of celestial compact objects in various gravity".

\section*{Appendix A: Deriving the analytic coupling constant at one-loop
approximation}

Using the K\"{a}ll\'{e}n-Lehman spectral representation, the analytic
coupling constant ($\alpha _{an}$) is defined as \cite{Shirkov1997}
\begin{equation}
\alpha _{APT}=\frac{1}{\pi }\int_{0}^{\infty }d\sigma \frac{\rho (\sigma ,a)%
}{\sigma +Q^{2}-i\epsilon},  \label{alpha-an}
\end{equation}%
where $\rho (\sigma ,a)=Im(\alpha _{RG}(-\sigma -i\epsilon ,a))$ is the
spectral density function calculated by the imaginary part of $\alpha _{RG}$
(the running coupling constant obtained from the renormalization group
equations). The spectral density function at one-loop approximation is given
by
\begin{equation}
\rho (\sigma ,a)=\frac{a^{2}\beta _{0}\pi }{\left( 1+a\beta _{0}\ln \sigma
/\mu ^{2}\right) ^{2}+(a\beta _{0}\pi )^{2}}.  \label{rho2}
\end{equation}

By inserting Eq. (\ref{rho2}) in Eq. (\ref{alpha-an}), we get

\begin{equation}
\alpha _{APT}^{(1)}=\frac{ 4\pi }{\beta _{0}}\left[\left( \ln \left( \frac{%
Q^{2}}{\Lambda ^{2}}\right) \right) ^{-1}+\frac{\Lambda ^{2}}{\Lambda
^{2}-Q^{2}}\right].
\end{equation}

\section*{Appendix B: The running coupling constant in BPT}

\begin{figure}[h]
	\center{\includegraphics[width=8.5cm] {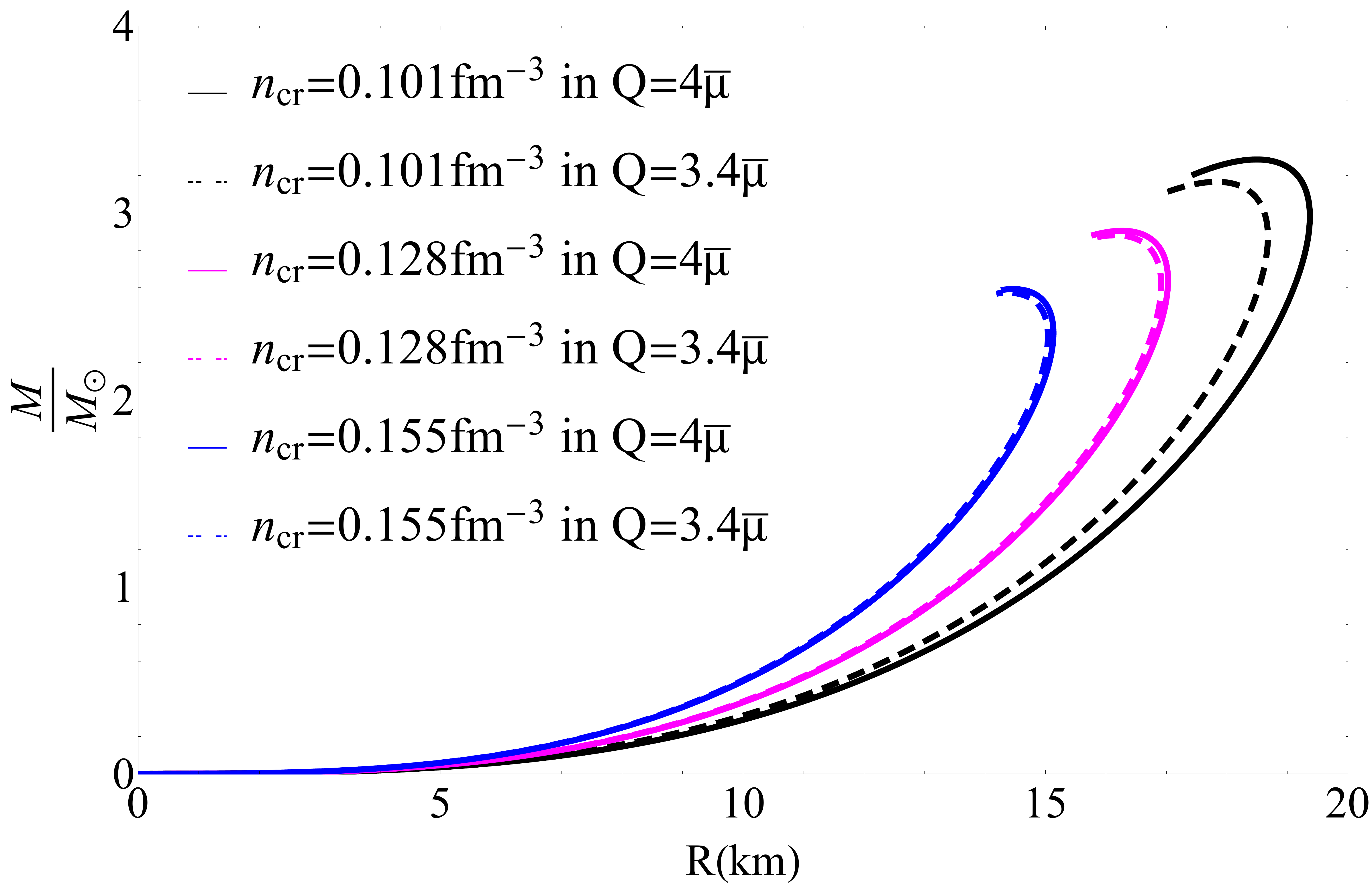}}
	\caption{\textcolor{black}{Mass versus radius of SQS in APT model for different choices of $Q$. Continuous and dashed lines correspond to $Q=4\overline{\mu}$ and $Q=3.4\overline{\mu}$,
		respectively. Each color indicates a $n_{cr}$ as follows. Continuous and
		dashed black lines correspond to $n_{cr}=0.101fm^{-3}$, continuous and
		dashed magenta lines correspond to $n_{cr}= 0.128fm^{-3}$, and continuous
		and dashed blue lines correspond to $n_{cr}= 0.155fm^{-3}$.}}
	\label{mass-comparing}
\end{figure}

The running coupling constant derived from BPT \cite%
{Simonov1995,Badalian1997,Simonov2011} is based on the success of static
potential $q\overline{q}$ in which the QCD interaction is divided into two
parts as $V(r)=-\frac{4\alpha _{V}}{3r}+\sigma r$. In this equation, the
first term is the perturbative part for short-distance interactions and the
second term is the non-perturbative part for the long-distance interactions.
In the BPT model, the gluon field is divided into perturbative and
non-perturbative parts. The non-perturbative part is a background field
which is described by the QCD string tension ($\sigma $). To avoid the $%
\Lambda $ pole problem, an effective gluonic mass ($m_{2g}^{2}=2\pi \sigma $%
) is assumed to be produced by this non-perturbative background field \cite%
{Simonov2011}. The IR behavior of the running coupling constant is then
modified by adding this non-perturbative part to all gluonic logarithms ($%
\ln Q^{2}\longmapsto \ln (Q^{2}+m_{2g}^{2})$) at all loop orders. At
one-loop approximation, the following equation is obtained
\begin{equation*}
\alpha _{BPT}^{(1)}(Q^{2})=\frac{4\pi }{\beta _{0}}\left( \ln \left( \frac{%
Q^{2}+m_{2g}^{2}}{\Lambda ^{2}}\right) \right) ^{-1},
\end{equation*}%
where, $\Lambda$ is $0.480$$GeV$.
\textcolor{black}{\section*{Appendix C: Structural properties of SQS in APT model for different values of renormalization scale.} As we mentioned in the introduction section, at zero temperature the phenomenological models suggest $Q=[1\overline{\mu}-4\overline{\mu}] (\overline{\mu}\equiv\sum_{f}\mu _{f}/3)$. We set the maximum value for $Q$ ($Q=4\overline{\mu}$) to find the maximum possible mass of the SQS in our models and  minimize the confinement effects. However we found that for $Q\gtrsim3.4\overline{\mu}$ the results do not change considerably, especially for $n_{cr}= 0.128fm^{-3}$ and $n_{cr}= 0.155fm^{-3}$. Figure. \ref{mass-comparing} and Table. \ref{different-Qs} show this feature for two different values of $Q$, including $Q=3.4\overline{\mu}$ and $Q=4\overline{\mu}$. We have defined $\delta_M\equiv\frac{M_{4\overline{\mu}}-M_{3.4\overline{\mu}}}{(M_{4\overline{\mu}}+M_{3.4\overline{\mu}})/2}$ and $\delta_R\equiv\frac{R_{4\overline{\mu}}-R_{3.4\overline{\mu}}}{(R_{4\overline{\mu}}+R_{3.4\overline{\mu}})/2}$ as a measure to see the difference of structural features in two different renormalization scales. As the Table. \ref{different-Qs} shows, the values of $\delta_M$ and $\delta_R$ are less than $5\%$. Specifically, the values of these quantities for $n_{cr}= 0.128fm^{-3}$ and $n_{cr}= 0.155fm^{-3}$ are negligible.}
\begin{table}[h]
	\small\addtolength{\tabcolsep}{-.5pt}
	\caption{\textcolor{black}{Structural properties of SQS for different values of $Q$}}
	\begin{tabular}{cccccccccc}
		\noalign{\hrule\vskip 1pt} 
		\noalign{\hrule\vskip 1pt}
		&
		\multicolumn{2}{c}{$Q/\overline{\mu}=3.4$}&
		\multicolumn{2}{c}{$Q/\overline{\mu}=4$} &
		\multicolumn{2}{c}{\%}
		\cr		
		\omit&\multispan2\hrulefill\hspace{0.1cm}&
		\multispan2\hrulefill\hspace{0.1cm}&
		\multispan2\hrulefill\hspace{0.1cm}&
		\cr
		\noalign{\vskip 1pt} 		
		$n_{cr}(fm^{-3})$ & $ M(M_{\odot}) $ & $R(km)$ & $ M(M_{\odot})$ & $R(km)$ & $\delta_M$ &$\delta_R$  
		\vspace*{.5mm} 
		\\
		\noalign{\hrule\vskip 1pt} 		
		\noalign{\hrule\vskip 1pt} 						
		$0.101$ & $ 3.16 $ & $17.74$ & $ 3.28$ & $18.52$ & $3.73$ & $4.30$\\
		$0.28$ & $ 2.88 $ & $16.14$ & $ 2.90$ & $16.80$ & $0.69$ & $2.41$\\
		$0.155$ & $ 2.57 $ & $14.45$ & $ 2.59$ & $14.52$ & $0.77$ & $0.48$\\
		\hline			
	\end{tabular}
	\label{different-Qs}
\end{table}
\end{document}